\documentclass[prl,twocolumn,showpacs,amsmath,amssymb]{revtex4}

\newcommand{\be}{\begin{equation}}
\newcommand{\ee}{\end{equation}}
\newcommand{\ba}{\begin{eqnarray}}
\newcommand{\ea}{\end{eqnarray}}
\newcommand{\ri}{\mathrm{i}}

\usepackage{graphicx}
\usepackage{dcolumn}

\begin{document}
  
\title{Nonlinear rupture of thin liquid films on solid surfaces}
\author{A. M. Leshansky$^1$}\email{lisha@caltech.edu} \author{B.Y. 
Rubinstein$^2$} 

\affiliation{$^1$Division of Chemistry $\&$ Chemical Engineering, California 
Institute of Technology, \\
Pasadena, CA 91125, USA \\ 
$^2$Department of Mathematics, University of California, \\
Davis, CA 95616, USA}

\date{\today}

\begin{abstract}
In this Letter we investigate the rupture instability of thin liquid films by means of a bifurcation analysis in the vicinity of the \emph{short-scale} instability threshold. The rupture time estimate obtained in closed form as a function of the relevant dimensionless groups is in striking agreement with the results of the numerical simulations of the original nonlinear evolution equations. This suggests that the weakly nonlinear theory captures the adequate physics of the instability. When antagonistic (attractive/repulsive) molecular forces are considered, nonlinear saturation of the instability becomes possible. We show that the stability boundaries are determined by the van der Waals potential alone. 
\end{abstract}


\maketitle

It is well known that a liquid film on a planar solid surface may become unstable due to long-range molecular forces. The forces originating from van der Waals attractions \cite{deryag55} accelerate thinning in  regions of film depression leading to film rupture and ``spinodal dewetting" \cite{rj74}. On the other hand, electrical double layers on the solid surface may give rise to intermolecular repulsions stabilizing thin films against rupture \cite{overbeek60}. 

In recent years, much effort has been put into theoretical modelling the dewetting phenomena \cite{wd82, oron97, bbd88, dewit94, ed93, rl00, zhang03, sharma03, ks01}.
A nonlinear theory of the film evolution based on the long-wave nature of the response was first posed in Ref. \onlinecite{wd82}. This approach, which has already been considered for different situations \cite{oron97}, yields nonlinear partial differential equations that describes the evolution of the interface shape, surfactant concentration,  \textit{etc.} Linear stability analysis is routinely applied to predict the onset of the instability and the characteristic wavelength, but the rupture time estimate obtained from the linear theory turns out to be rather poor: it underestimates the rupture time due to highly nonlinear nature of response. The most common and straightforward approach is to solve the evolution equations numerically \cite{bbd88}, \cite{dewit94} \cite{zhang03}, \cite{becker03,ks01,sharma03,shj01}. The obvious disadvantage of the numerical simulation is that for a complex problem that involves many parameters, full parametric study of the rupture is quite elaborate. 

A bifurcation technique was first applied in \cite{ed93} to arrive at the nonlinear estimate for the rupture time in the vicinity of a steady bifurcation point. It was demonstrated that nonlinear terms owing to van 
der Waals attractions contribute to rapid acceleration of the rupture beyond the 
linear regime. Analysis of the nonlinear evolution of small disturbances leads 
to a dynamic Landau equation for the perturbation amplitude. The closed form
solution of the amplitude equation provides a time for ``blowup" of the initial 
small-amplitude disturbance that was proposed to be a good estimate of the 
nonlinear rupture time. The approach has never been given enough attention 
perhaps because the analysis involves rather tedious algebra and can only 
be done ``by hand" for some simple cases. It has been demonstrated 
in \cite{rl00} that the derivation of the amplitude equation can be
automatized by using a previously developed symbolic algorithm for bifurcation 
analysis \cite{rp99}. Although the closed form nonlinear estimate of the rupture 
time of the thin film in presence of insoluble surfactant was derived in
\cite{rl00}, the lack of parametric study of the problem by simulations didn't 
allow a proper comparison of the two approaches. Recently, an extensive 
numerical study of the thin film rupture driven by van der Waals attractive 
forces in the presence of insoluble surfactant and hydrodynamic slip was reported in \cite{zhang03}. We have developed a generalized theory of thin film rupture for an arbitrary intermolecular potential; further, we compare the rupture time estimate from our theory with the results of simulations by \cite{zhang03} for the purely attractive potential and come up with some predictions for the competing (attractive/repulsive) potential.

We consider a model describing the evolution of a thin liquid film a solid substrate subject to a van der Waals force in the presence of a slip and insoluble surfactant. The dimensionless film thickness $h$ and surfactant concentration $\Gamma$ are governed by a system of coupled evolution equations derived
in the long-wave approximation
\cite{zhang03},
\ba
h_t&=&\left[\mathcal{M}\Gamma_x h \left(\frac{h}{2}+\beta\right)-\mathcal{F}_x 
h^2\left(\frac{h}{3}+\beta\right) \right]_x, \label{eq-h}\\
\Gamma_t&=&\frac{\Gamma_{xx}}{\mathcal{P}}+
\left[\mathcal{M}\Gamma\Gamma_x(\beta+h)-\Gamma \mathcal{F}_x
h\left(\frac{h}{2}+\beta\right) \right]_x, \label{eq-Gamma}
\ea
with $$\mathcal{F}(x,t)=-\varphi(h)+{\mathcal C}h_{xx}$$ 
and where $\mathcal{M}$ is a Marangoni number, $\mathcal{P}$ is a Peclet number, $\beta$ is a Navier slip coefficient, $\mathcal{C}$ is a surface tension parameter and 
$\varphi=(\partial \Delta G/\partial h)$ is the van der Waals potential (all dimensionless). For nonslipping films ($\beta=0$) equations (\ref{eq-h}-\ref{eq-Gamma}) are equivalent to those in \cite{dewit94}.

The linear stability analysis of the uniform stationary state ${\bf u}_0=\left\{h_0,\Gamma_0 \right\}$  results 
in a critical value of the wavenumber corresponding to a stationary
bifurcation point $k_{c}=(-\varphi'(h_0)/{\cal C})^{1/2}$. 
Following a standard procedure we choose $\mathcal{C}$ as a bifurcation parameter. In a bounded domain, $0<x<\mathcal{L}$, the basic solution ${\bf u}_0$ changes stability (becomes spinodally unstable) with $k=k_c$ and when $\mathcal{C}<\mathcal{C}_{c}=-\varphi'(h_0)\mathcal{L}^2/4\pi^2$, where $\mathcal{C}_{c}$ correspond to a steady bifurcation point. 

To investigate the nonlinear problem in the vicinity of the bifurcation point we expand the bifurcation parameter as 
$\mathcal{C}=\mathcal{C}_c+\epsilon^2\mathcal{C}_2+...$, where $\epsilon$ is a small criticality, introduce a slow time scale suggested by the linear theory, $\tau=\epsilon^2 t$, and seek the solution in power series of $\epsilon$ as ${\bf u}={\bf u}_0+\epsilon{\bf u}_1+...$. Substitution of this expansion into the system (\ref{eq-h}-\ref{eq-Gamma}) to the first order in $\epsilon$ yields ${\bf u}_1=\left(A(\tau) e^{\ri k_c x}+\mbox{c.c.}\right)\:{\bf U}$, where ${\bf U}=\{1,0\}$ is a solution of the linearized zero-eigenvalue problem.  The complex amplitude $A$ satisfies the dynamic Landau equation which is determined to $\mathcal{O}(\epsilon^3)$ of the perturbation theory:
\be
\frac{\partial A}{\partial \tau}=\alpha A+\kappa |A|^2\:A\:, \label{GL}
\ee
where the linear coefficient $\alpha$ and the Landau coefficient $\kappa$ are given by 
\be
\alpha=- \delta\: k_c^4\:\mathcal{C}_2, \qquad \kappa=
\frac{\delta}{6\:\mathcal{C}_c}\: \left(\varphi''^2+3\varphi'\varphi'''\right), \label{cubic}
\ee
respectively, and
$$
\delta=\frac{h_0^2\left[4(h_0+3\beta)+h_0(h_0+4\beta)\theta\right]}{12\:\left[1+(h_0+\beta)\theta\right]} , \ \ \theta=\mathcal{M}\mathcal{P}\Gamma_0. \ \ 
$$

The closed form solution of the amplitude equation (\ref{GL}) can be easily obtained given the 
initial value of the amplitude $A_0=A(0)$.
The ''blowup" time, corresponding to the infinite growth of the amplitude $A$ 
and providing a nonlinear estimate of the rupture time (in the original time scale),
in the vicinity of the bifurcation point, as $\mathcal{C}_2\rightarrow0$ is determined 
solely by the cubic coefficient
\be
t_{rup} \approx (2 A_0^2 \kappa)^{-1}\:, \label{trup2}
\ee
where $A_0$ is now $\mathcal{O}(\epsilon)$.
An important observation is that despite the complicated nature of the original evolution equations, the structure of the cubic coefficient $\kappa$ in (\ref{cubic}) is very simple: terms in brackets contain only derivatives of the intermolecular potential $\varphi$ and a factor $\delta/\mathcal{C}_c$ incorporates the dependence on the rest of parameters. Since $\delta>0$, it is readily seen from (\ref{cubic}) that for purely attractive potential $\kappa$ is always positive and the rupture is inevitable. For the most commonly encountered attractive potential $\phi=\mathcal{A}/h^c$, with $c=3,4$ (unretarded and retarded case, respectively) we calculate the rupture time from (\ref{trup2}) and compare to the results of numerical simulations of the original evolution equations (\ref{eq-h}-\ref{eq-Gamma}) reported in \cite{zhang03}. $\mathcal{A}=\mathcal{A}_*/6\pi\rho\nu^2h_*^{c-2}$ is the scaled Hamaker constant with $h_*$ being the mean film thickness (here and thereafter dimensional quantities are marked with ${}_*$). 
Typical evolution of the  film thickness in numerical simulations shows an accelerated thinning of the film in the depressed region due to the initial disturbance at some unstable wavelength $\lambda>2\pi/k_c$ with and subsequent film rupture. This behavior suggests that the acceleration of the film rupture is due to a nonlinear self-coupling of the perturbation beyond the linear regime and therefore the cubic nonlinearity in the amplitude equation (\ref{GL}) should provide an accurate description of the nonlinear rupture. Since (\ref{trup2}) is formally valid in the close vicinity of the instability threshold, it is considered as a nonlinear approximation for the rupture time, $t_{rup}$, far from the instability threshold (as in \cite{zhang03}) with one adjustable parameter, $A_0$.  
\begin{figure}
\hspace{0.2cm}\includegraphics[width=3.2in, height=2.4in]{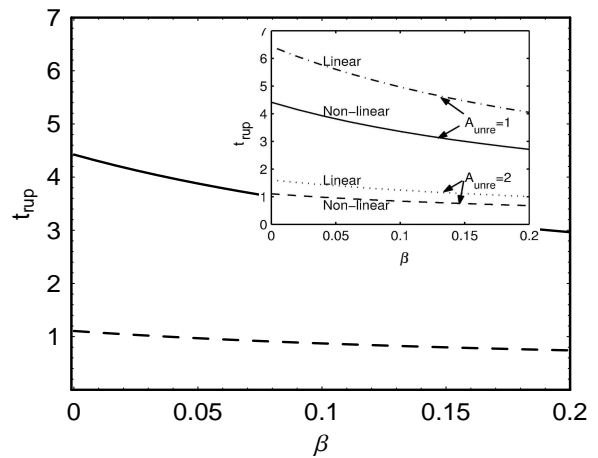}
\caption{\label{fig1} Variation of rupture time vs. $\beta$ with $A_0=0.106$, $h_0=1$, $\mathcal{C}=1$, $\mathcal{M}=1$, $\mathcal{P}=100$, $\Gamma_0=0.5$, $c=3$ and ${\cal A}=1$ (\hbox{---$\!$---}), ${\cal A}=2$ (\hbox{{--}\,{--}\,{--}}). The inset shows analogous results of the numerical simulations \cite{elsevier}.}
\end{figure}
\begin{figure}
\includegraphics[width=3.2in, height=2.4in]{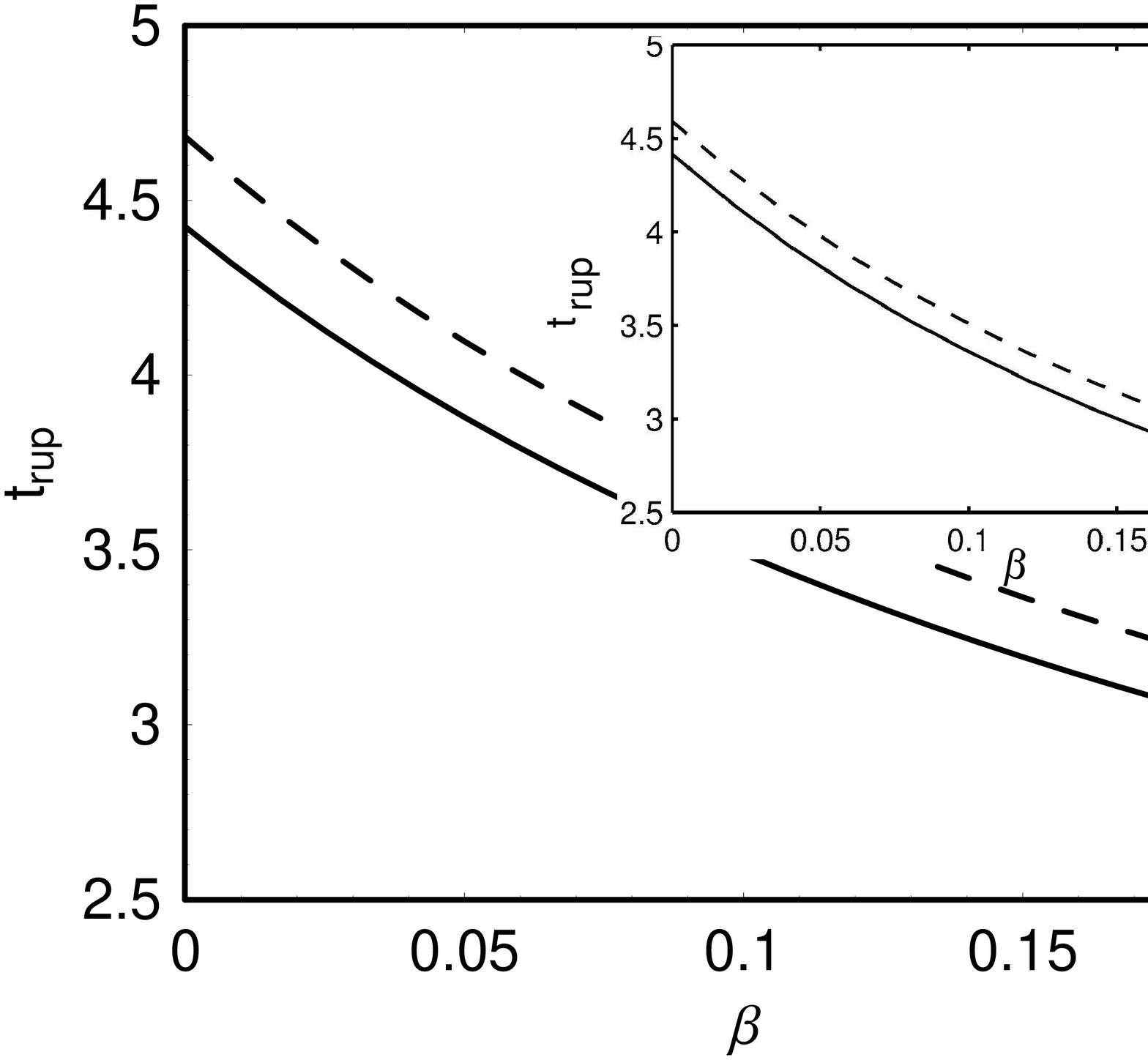}
\caption{\label{fig2} Variation of rupture time with $\beta$ for ${\cal A}=1$ and the same values of other parameters as in Fig. \ref{fig1} except we vary ${\cal M}$: ${\cal M}=1$ (\hbox{---$\!$---}), ${\cal M}=100$ (\hbox{{--}\,{--}\,{--}}). The inset shows analogous results of the numerical simulations \cite{elsevier}.}
\end{figure}
\begin{figure}
\hspace{0.2cm}
\includegraphics[width=3.0in, height=2.2in]{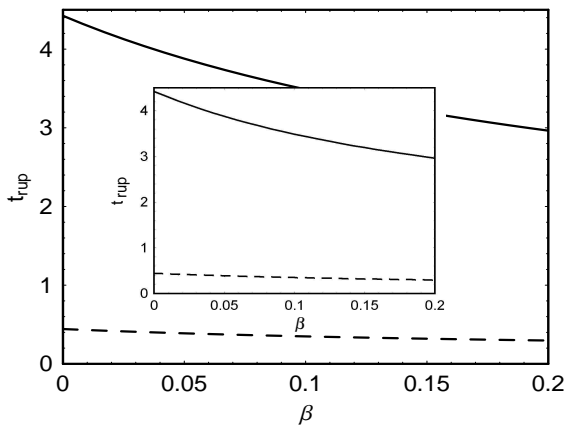}
\caption{\label{fig3} Variation of rupture time with $\beta$ for ${\cal A}=1$ and the same values of other parameters as in Fig. \ref{fig1} except now we vary ${\cal C}$: ${\cal C}=1$ (\hbox{---$\!$---}); ${\cal C}=0.1$ (\hbox{{--}\,{--}\,{--}}). The inset shows analogous results of the numerical simulations \cite{elsevier}.}
\end{figure}
Figures \ref{fig1}--\ref{fig3} show the comparison between the theory with $A_0=0.106$ and the simulations \cite{zhang03}. It is readily seen that there is an excellent \textit{quantitative} agreement with the numerical estimate of $t_{rup}$ as a function of different dimensionless parameters.  
An obvious advantage of the present analysis is that the closed form expression for the 
nonlinear rupture time as a function of different parameters of the problem is 
available in a closed form expression for a general van der Waals potential.

When the antagonistic attractive/repusive intermolecular interactions are present, nonlinear saturation of the 
rupture instability is possible as $\kappa$ may change sign (supercritical bifurcation). Let us consider the general representation of the antagonistic van der Waals potential \cite{ks01,sharma03}
\be
\varphi(h)=\frac{\mathcal{A}}{h^c}-\frac{\mathcal{B}}{h^d}
-(S_p/l_p) \exp(-h/l_p), \label{def-phi}
\ee
where $\mathcal{A}$ is defined as before, $\mathcal{B}=\mathcal{B}_*/\rho\nu^2h_*^{d-2}$,
$S_p={S_{p}}_*h_*^2/\rho\nu^2$ and $l_p={l_p}_*/h_*$. It follows from (\ref{cubic}) that the stationary nonruptured solution with amplitude $A=(-\alpha/\kappa)^{1/2}$ is stable if
\be
d\gamma_1 + H\gamma_2 <  c \:,
\qquad \sum_{i,j;}\sum_{i+j\le 2} a_{ij} \gamma_1^i \gamma_2^j <  0 \:, \label{gamma12}
\ee
where the first inequality is imposed by the linear theory,
$\gamma_1={\cal B} h_0^{c-d}/{\cal A}$, $\gamma_2 = (S_p/l_p) h_0^{c} e^{-H}/{\cal A}$, $H=h_0/l_p$
and $a_{ij}$ are some polynomial functions of $c,d$ and $H$ only.

When the interplay between algebraic potentials is considered ($\gamma_2=0$), the nonlinear stability region is defined by $\gamma_1$ alone. For exponents $(c,d)=(3,4)$ (repulsive retarded van der Waals force) the film is stable if $0.51<\gamma_1<0.73$.
\begin{figure}
\centering{\includegraphics[width=2.8in, height=2.1in]{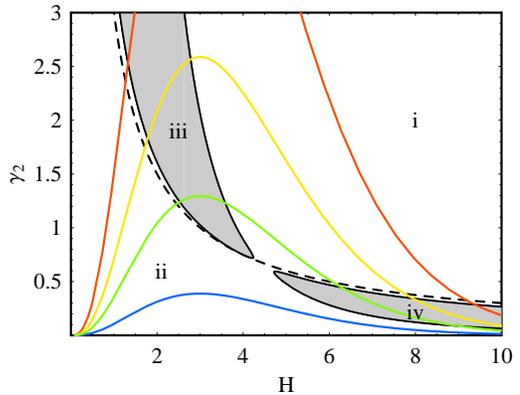}}
\caption{\label{fig4} Stability diagram for an antagonistic 
exponential/algebraic potential ($\gamma_1=0$) in plane of parameters $\gamma_2=(S_p/l_p) h_0^{c} e^{-h_0/l_p}/{\cal A}$ and $H=h_0/l_p$. The dashed curve corresponds to marginal stability boundary with $c=3$. The regions of stationary nonruptured states are shown in gray. Colored curves show the dependence $\gamma_2$ vs. $H$ for varying hydrophobicity of the substrate from hydrophilic (blue curve) to hydrophobic (red curve) using parameters from \cite{ks01}.}
\end{figure}
\begin{figure}
\centering{\includegraphics[width=2.8in, height=2.1in]{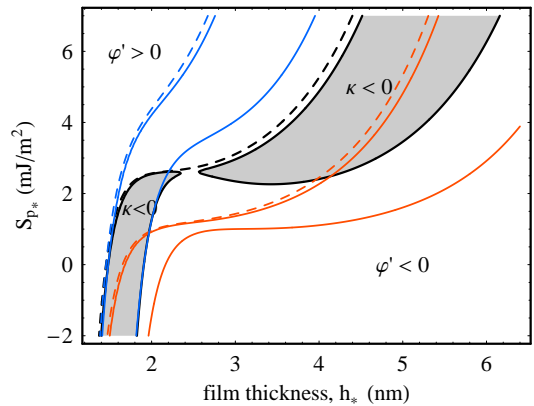}}
\caption{\label{fig5} Stability diagram for a general van der Waals potential (\ref{def-phi}). The regions right to the dashed curves are spinodally unstable and the regions between the solid curves correspond to stationary nonruptured state. Black curves correspond to $\mathcal{A}_*=3.0\times10^{-20}$ J, $\mathcal{B}_*=5.04\times10^{-75}$~J~m$^6$ and ${l_p}_*=0.6$ nm (nonlinear stability regions are shown in gray). The color curves correspond to the same values of the parameters except ${l_p}_*=0.4$~nm (blue curves) and $A=1.4\times10^{-20}$~J (red curves).}
\end{figure}
For the exponents $(3,9)$ (short-range Born repulsion) the stability window is shifted to lower values of $\gamma_1$ and the film is stable whenever $0.066<\gamma_1<0.30$. For instance, using the values of the Hamaker constants measured for a polystyrene film on oxidized Si wafers with $\mathcal{A}_*=2.2\times10^{-20}$\ J  and $\mathcal{B}_*=5.04\times10^{-75}$~J~m$^6$ \cite{becker03} the nonlinear analysis predicts a stable film thickness of $h_0\approx2$~nm ($\gamma_1=0.068$) while the linear theory results in $1.56$~nm and the equilibrium thickness determined from the minimum of $\Delta G$ is only $1.3$~nm \cite{becker03}. For thicker films $\gamma_1$ is rapidly decreasing as $h_0^{-6}$, \textit{e.g.} $\gamma_1\approx 0.00125$ for $h_0=3.9$~nm. Although, a qualitative difference in the morphology of dewetting in \cite{becker03} was observed for thicker films ($3.9$ nm vs. $4.9$ nm) we speculate that the qualitative change is due to approach to the nonlinear stability boundary as $h_*$ approaches the thickness of $2$ nm.

When the interplay between algebraic and exponential potentials is considered ($\gamma_1=0$), the nonlinear stability diagram can be defined in terms of $\gamma_2$ and $H$ as in Fig. \ref{fig4}.  The dashed lines correspond to the boundary of linear stability, while the regions of nonlinear stability corresponding to $\kappa<0$ are shown in gray (\textrm{iii} and \textrm{iv}). Note that, if the long-range attraction is combined with the shorter-range repulsion ($\mathcal{A}>0$, $S_p>0$), the spinodally unstable region is under the dashed curve (region \textrm{ii} in Fig. \ref{fig4}), and nonlinear saturation occurs for $H \gtrsim 4.5$ and small values of $\gamma_2$ (region \textrm{iv}). In the opposite case ($\mathcal{A}<0$, $S_p<0$) the spinodally unstable region lies above the dashed curve (region \textrm{i}) and stabilization occurs for thinner films, $H\lesssim 4$, and moderate  values of $\gamma_2$ (region \textrm{iii}). For instance, for aqueous films on \textrm{Si} substrates with $c=3$, $\mathcal{A}_*=-1.41\times10^{-20}$~J and ${l_p}_*=0.6$~nm \cite{ks01} we plot $\gamma_2$ vs. $H$ in Fig. \ref{fig4} (color curves) for different values of ${S_p}_*$ varying from $-0.61$~mJ/m$^2$ (blue) to $-8.5$~mJ/m$^2$(red) due to increasing hydrophobicity of the substrate \cite{ks01}. It is evident from Fig. \ref{fig4} that the emergence of stationary nonruptured ultrathin films is possible on non-hydrophilic substrates as the color curves cross region (\textrm{iii}), while on hydrophilic substrates (the blue curve) the film of any thickness is stable, in accord with \cite{ks01}.

More interesting behavior is anticipated for competing short-range algebraic and exponential potentials. In this case $\gamma_1,\gamma_2 \ne 0$ and as they both vary with $h_0$ we chose to depict the stability diagram in terms of dimensional quantities, ${S_p}_*$ and $h_*$ as in Fig. \ref{fig5}. It is evident that stabilization is possible for a wide range of film thicknesses, $h_*$. When the magnitude of the exponential repulsion is small, the steady nonruptured state is only possible for ultrathin films;  for moderate values of ${S_p}_*$ the band of stable solutions widens. For instance, when ${S_p}_*=1.1$~mJ/m$^2$, with parameters corresponding to the red curve in Fig. \ref{fig5}, nonlinear theory predicts that the film is stable below a thickness of $\sim 4$~nm, whereas linear stability provides a value of $\sim2.2$~nm. 

Finally, we consider van der Waals interactions of the polystyrene films with \textrm{SiO} coating on \textrm{Si} substrate (in dimensional form) $\varphi_*=\frac{\mathcal{A}_{*\textrm{Si}}-\mathcal{A}_{*\textrm{SiO}}}{6\pi(h_*+d_*)^3}+\frac{\mathcal{A}_{*\textrm{SiO}}}{6\pi h_*^3}-\frac{\mathcal{B}_*}{h_*^9}$ with $\mathcal{A}_{*\textrm{Si}}=-1.3\times10^{-19}$~J, $A_{*\textrm{SiO}}=2.2\times10^{-20}$~J and $\mathcal{B}_*=5.04\times10^{-75}$~J~m$^6$  \cite{shj01}. In this case the stability diagram can be depicted in terms of the film thickness, $h_*$, and the \textrm{SiO} coating thickness, $d_*$ (not shown). Again, there is a narrow stability window for ultrathin films up to $\sim 2$~nm while its width is insensitive to the variation in \textrm{SiO} coating thickness. The stabilization for thicker films does not materialize similar to the previously discussed case of exponents $(3,9)$ without coating.

In conclusion, we have developed a nonlinear theory for the rupture of a thin liquid film subject to a general van der Waals potential. The comparison between the prediction of the weakly nonlinear analysis and the numerical results is provided for the first time; it is demonstrated that there is an excellent \textit{quantitative} agreement between the nonlinear rupture time estimate from our theory and the numerical estimate. When an antagonistic potential is considered, the saturation of the rupture instability beyond linear regime is possible, while the stability boundary is determined solely by the intermolecular potential. The results concerning the existence of steady nonruptured  states should be accessible via numerical simulations. 




\end{document}